\input lanlmac
\input epsf.tex
\input mssymb.tex
\overfullrule=0pt

\newcount\figno
\figno=0
\def\fig#1#2#3{
\par\begingroup\parindent=0pt\leftskip=1cm\rightskip=1cm\parindent=0pt
\baselineskip=11pt
\global\advance\figno by 1
\midinsert
\epsfxsize=#3
\centerline{\epsfbox{#2}}
\vskip 12pt
{\bf Fig.\ \the\figno:} #1\par
\endinsert\endgroup\par
}
\def\figlabel#1{\xdef#1{\the\figno}%
\writedef{#1\leftbracket \the\figno}%
}
\def\omit#1{}

\def\pre#1{{\tt
#1}}

\def\Rc{{\check R}}

\def\der{\partial}

\def\ket#1{\left| #1 \right>}
\def\qed{\nobreak\hfill\vbox{\hrule height.4pt%
\hbox{\vrule width.4pt height3pt \kern3pt\vrule width.4pt}\hrule height.4pt}\medskip\goodbreak}
\lref\Lu{G. Luzstig, {\sl Canonical bases arising from quantized enveloping algebras}, {\it J. Amer. Math. Soc.} 3 (1990), 447--498.}
\lref\RS{A.V. Razumov and Yu.G. Stroganov, 
{\sl Combinatorial nature
of ground state vector of $O(1)$ loop model},
{\it Theor. Math. Phys.} 
{\bf 138} (2004) 333-337; {\it Teor. Mat. Fiz.} 138 (2004) 395-400, \pre{math.CO/0104216}.}
\lref\BdGN{M.T. Batchelor, J. de Gier and B. Nienhuis,
{\sl The quantum symmetric XXZ chain at $\Delta=-1/2$, alternating sign matrices and 
plane partitions},
{\it J. Phys.} A34 (2001) L265--L270,
\pre{cond-mat/0101385}.}
\lref\DFZJ{P.~Di Francesco and P.~Zinn-Justin, {\sl Around the Razumov--Stroganov conjecture:
proof of a multi-parameter sum rule}, {\it E. J. Combi.} 12 (1) (2005), R6,
\pre{math-ph/0410061}.}
\lref\DFZJb{P.~Di Francesco and P.~Zinn-Justin, {\sl Inhomogeneous model of crossing loops
and multidegrees of some algebraic varieties}, to appear in {\it Commun. Math. Phys} (2005),\break
\pre{math-ph/0412031}.}
\lref\KZJ{A. Knutson and P. Zinn-Justin, {\sl A scheme related to the Brauer loop model}, \pre{math.AG/0503224}.}
\lref\Pas{V.~Pasquier, {\sl Quantum incompressibility and Razumov Stroganov type conjectures},
\pre{cond-mat/0506075}.}
\lref\Jo{A. Joseph, {\sl On the variety of a highest weight module}, {\it J. Algebra} 88 (1) (1984), 238--278.}
\lref\Ro{W. Rossmann,
{\sl Equivariant multiplicities on complex varieties.
  Orbites unipotentes et repr\'esentations, III},
  {\it Ast\'erisque} No. 173--174, (1989), 11, 313--330.}
\lref\Ho{R.~Hotta, {\sl On Joseph's construction of Weyl group representations}, Tohoku Math. J. Vol. 36
(1984), 49--74.}
\lref\KZ{V.~Knizhnik and A. Zamolodchikov, {\sl Current algebra and Wess--Zumino model in two dimensions},
{\it Nucl. Phys.} B247 (1984), 83--103.}
\lref\FR{I.B.~Frenkel and N.~Reshetikhin, {\sl Quantum affine Algebras and Holonomic Difference Equations},
{\it Commun. Math. Phys.} 146 (1992), 1--60.}
\lref\JM{M.~Jimbo and T.~Miwa, {\it Algebraic analysis of Solvable Lattice Models}, 
CBMS Regional Conference Series in Mathematics vol. 85, American Mathematical Society, Providence, 1995.}
\lref\DFKZJ{P.~Di Francesco, A. Knutson and P. Zinn-Justin, work in progress.}
\lref\AS{F.~Alcaraz and Y.~Stroganov, {\it The Wave Functions for the Free-Fermion Part of the Spectrum
of the $SU_q(N)$ Quantum Spin Models}, \pre{cond-mat/0212475}.}
\Title{SPhT-T05/130}
{\vbox{
\centerline{Quantum Knizhnik--Zamolodchikov equation,}
\medskip
\centerline{generalized Razumov--Stroganov sum rules}
\medskip
\centerline{and extended Joseph polynomials}
}}
\bigskip\bigskip
\centerline{P.~Di~Francesco \footnote{${}^\#$}
{Service de Physique Th\'eorique de Saclay,
CEA/DSM/SPhT, URA 2306 du CNRS,
C.E.A.-Saclay, F-91191 Gif sur Yvette Cedex, France}
and P. Zinn-Justin \footnote{${}^\star$}
{Laboratoire de Physique Th\'eorique et Mod\`eles Statistiques, UMR 8626 du CNRS,
Universit\'e Paris-Sud, B\^atiment 100,  F-91405 Orsay Cedex, France}}
\vskip0.5cm
\noindent
We prove higher rank analogues of the Razumov--Stroganov sum rule for
the groundstate of the $O(1)$ loop model on a semi-infinite cylinder: we show that
a weighted sum of components of the groundstate of the $A_{k-1}$ IRF model
yields integers that generalize the numbers of alternating sign matrices.
This is done by constructing minimal polynomial solutions of the level 1
$U_q(\widehat{\goth{sl}(k)})$ quantum Knizhnik--Zamolodchikov equations, which may
also be interpreted as quantum incompressible $q$-deformations of fractional
quantum Hall effect wave functions at filling fraction $\nu=1/k$. In addition
to the generalized Razumov--Stroganov point $q=-e^{i\pi/k+1}$, another 
combinatorially interesting point is reached in the rational limit $q\to -1$,
where we identify the solution with extended Joseph polynomials associated
to the geometry of upper triangular matrices with vanishing $k$-th power.

\bigskip

\def\cR{{\check R}}

\Date{08/2005}
%
%
\newsec{Introduction}
The present work stems from
the recent activity around the so-called Razumov--Stroganov conjecture 
\refs{\BdGN,\RS},
which relates 
the groundstate vector of the $O(1)$ loop model on a semi-infinite cylinder of perimeter $2n$
to the numbers of configurations
of the six--vertex (6V) model on a square grid of size $n\times n$, with domain wall boundary conditions (DWBC). In
\DFZJ, we were able
to prove a weaker version of the conjecture, 
identifying the {\it total number}\/ of configurations,
also equal to the celebrated number of $n\times n$ alternating sign matrices,
to the {\it sum}\/ of entries of the groundstate vector, 
thus establishing the Razumov--Stroganov sum rule.
The proof is more general and actually identifies the groundstate $\Psi$
of the fully inhomogeneous version of the loop model with the so-called Izergin--Korepin determinant, i.e.
the inhomogeneous partition function of the 6V model with DWBC. It makes extensive use of the integrability
of the models. We adapted this proof to the case of the $O(1)$ crossing loop model \DFZJb, 
which is based on a rational 
(as opposed to trigonometric) solution of the Yang--Baxter equation,
and established a direct relation 
to the geometry of certain schemes of matrices,
further developed 
in \KZJ. 

More recently, Pasquier \Pas\ constructed a polynomial representation of the affine Hecke algebra, 
allowing him to recover the aforementioned sum rule.
His method can be reformulated as finding a polynomial solution $\Psi$ to
the quantum Knizhnik--Zamolodchikov ($q$KZ) equation for $U_q(\widehat{\goth{sl}(2)})$ at level 1. 
At $q=-e^{i\pi/3}$
it coincides with the Razumov--Stroganov groundstate, but it is 
defined for arbitrary $q$ as well, although it has no direct interpretation as the groundstate of some loop
model. 
We note here that for $q\to -1$,
when the trigonometric solution of the Yang--Baxter equation degenerates into a rational
one, the entries of $\Psi$ tend to non-negative integers 
(different from the Razumov--Stroganov case).
As we shall see below, these are the degrees of the components of
the scheme of 
upper triangular complex $2n\times 2n$ matrices with square zero. 

In this note, we also address the case of higher rank algebras, i.e.\ the so-called $A_{k-1}$
vertex/height models. 
The natural rank $k$
counterpart of the $O(1)$ loop model is constructed by using a path representation for the Hecke algebra 
quotient associated to $U_q(\goth{sl}(k))$ (we do not use here the more traditional spin chain representation,
see e.g.\ the somewhat related work \AS;
though we mention it a few times in what follows). We then work out the polynomial solution $\Psi$ to the
$U_q(\widehat{\goth{sl}(k)})$ $q$KZ equation at level 1, which, at the particular point $q=-e^{i\pi \over k+1}$,
turns out to be exactly the groundstate of a half-cylinder $A_{k-1}$ IRF model, with a transfer matrix 
acting naturally in the path representation. At this generalized ``Razumov--Stroganov" point,
we establish a new sum rule for the entries of $\Psi$. In particular, we find natural generalizations
of the number of alternating sign matrices for arbitrary $k$. What exactly is counted by these numbers
still eludes us.
We finally investigate the rational  $q\to -1$ limit within the framework of
equivariant cohomology in the space of upper triangular matrices, relating it
to schemes of nilpotent matrices of order $k$.

\newsec{$A_{k-1}$ $R$-matrix and path representation}
The standard abstract trigonometric solution of the Yang-Baxter equation reads
\eqn\trigosol{ \Rc_i(z,w)={q z -q^{-1} w\over q w-q^{-1} z} +{z-w\over q w-q^{-1} z}e_i }
where the $e_i$, $i=1,2,\ldots,N-1$ are the  generators of the Hecke algebra $H_N(\tau)$, with the relations
\eqn\relahec{ e_ie_j=e_je_i,\ |i-j|>1, \qquad e_ie_{i+1}e_i-e_i=e_{i+1}e_ie_{i+1}-e_{i+1},
\qquad e_i^2=\tau e_i}
with the parametrization $\tau=-(q+q^{-1})$.

In this note we restrict the $e_i$ to be the generators of the 
quotient of the Hecke algebra related to $U_q(\goth{sl}(k))$ in the fundamental
representation, and
denoted by $H_N^{(k)}(\tau)$; it is obtained by imposing
extra relations, namely the vanishing of the $q$-antisymmetrizers of order $k$, $Y_k(e_i,e_{i+1},\ldots,e_{i+k-1})$,
defined recursively by $Y_1(e_i)=e_i$, and 
$Y_{m+1}(e_i,\ldots,e_{i+m})=Y_m(e_i,\ldots,e_{i+m-1})(e_{i+k}-\mu_m)Y_m(e_i,\ldots,e_{i+m-1})$,
with $\mu_m=U_{m-1}(\tau)/U_{m}(\tau)$, $U_m$ the Chebyshev polynomials of the second kind 
$U_m(2\cos\theta)=\sin\big((m+1)\theta\big)/\sin(\theta)$.
For $k=2$, the quotient
$H_N^{(2)}(\tau)$ is nothing but the Temperley-Lieb algebra $TL_N(\tau)$.

Representations of the algebra $H_N^{(k)}(\tau)$ have
been extensively used to construct lattice integrable vertex models based on $A_{k-1}$.
Here, we consider the so-called path representation, naturally leading to IRF models,
and restrict ourselves to the case where $N=nk$.
States are indexed by closed paths from and to the origin
on the (link-oriented) Weyl chamber of $SU(k)$, allowed to
make steps $u_1,u_2,\ldots,u_k$, where $u_1=\omega_1$, $u_2=\omega_2-\omega_1$, \dots, $u_k=-\omega_{k-1}$
in terms of the fundamental weights $\omega_i$. These paths visit only points 
$\lambda=\sum_{1\leq i\leq k-1}\lambda_i\omega_i$ with all $\lambda_i\geq 0$. It is useful to represent
them via their sequence of steps, substituting $u_i\to i$ for simplicity. For instance,
the path of length $nk$
closest to the origin is $(12\ldots k)^n$, namely $n$ repetitions of the sequence $1,2,\ldots,k$,
and we denote it by $\pi_f$.
Likewise, the path farthest from the origin is $(1)^n(2)^n\ldots(k)^n$, and we denote it by $\pi_0$.
A useful notation consists in representing each step $j$ by a unit segment forming an angle of 
${\pi(k+2-2j)\over 2(k+1)}$ with the horizontal direction: each $\pi$ becomes a broken line touching
the $x$ axis at its ends and staying above it. There are exactly $(kn)!\prod_{0\leq j\leq k-1} j!/(n+j)!$
such paths. For illustration, for $k=3$, $n=2$ we have the five following $A_2$ paths of length $6$:
\eqn\fivepaths{\eqalign{
\pi_0&=(112233)\ \ \ \to \ \ \ \epsfxsize=1.5cm\vcenter{\hbox{\epsfbox{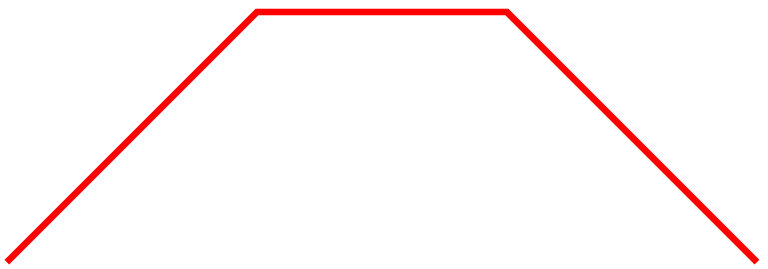}}}\qquad\qquad
\pi_1=(112323)\ \ \ \to \ \ \ \epsfxsize=1.5cm\vcenter{\hbox{\epsfbox{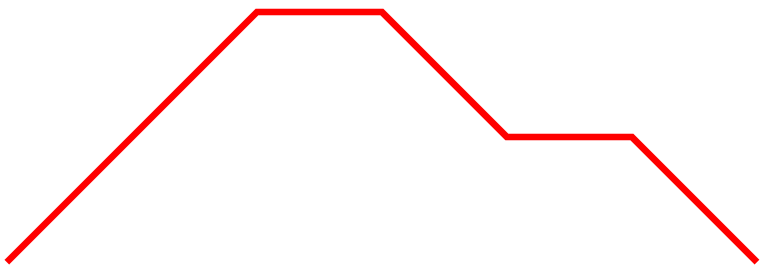}}}\cr
\pi_2&=(121233)\ \ \ \to \ \ \ \epsfxsize=1.5cm\vcenter{\hbox{\epsfbox{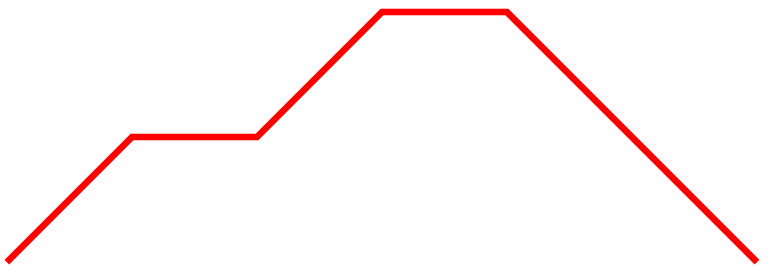}}}\qquad\qquad
\pi_3=(121323)\ \ \ \to \ \ \ \epsfxsize=1.5cm\vcenter{\hbox{\epsfbox{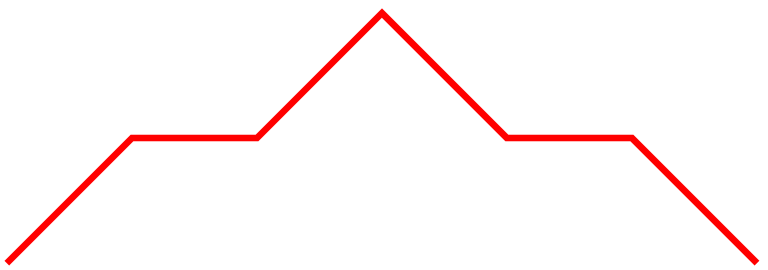}}}\cr
\pi_f&=(123123)\ \ \ \to \ \ \ \epsfxsize=1.5cm\vcenter{\hbox{\epsfbox{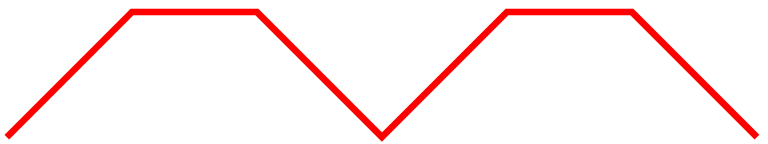}}}\cr}}
shown in step- and broken line- representations.

Associating a vector $\ket{\pi}\equiv
\ket{\pi_1 \pi_2 \ldots\pi_N}$ to each path, the path representation satisfies
the following properties:
\item{\bf (P1)} $e_i \ket{\pi} =\tau \ket{\pi}$ 
if $\pi_i<\pi_{i+1}$ ($\pi$ locally convex)
\item{\bf (P2)} $e_i \ket{\pi}  =\sum_{\pi'} C_{i,\pi,\pi'} \ket{\pi'}$
if $\pi_i\geq \pi_{i+1}$ ($\pi$ locally flat or concave), for some 
$C_{i,\pi,\pi'}\in \{0,1\}$

\noindent and will be discussed in detail elsewhere. Let us however
mention two more properties of crucial importance in what follows:
\item{\bf (P3)} If $C_{i,\pi,\pi'}=1$ then $\pi'$ 
is locally convex between steps 
$i$ and $i+1$, namely $\pi'_i<\pi'_{i+1}$
\item{\bf (P4)} If $C_{i,\pi,\pi'}=1$ then either $(\pi_i,\pi_{i+1})=(\pi_{i+1}',\pi_i')$
and $\pi_m=\pi'_m$ for all $m\neq i,i+1$, i.e. $\pi'$ exceeds $\pi$ by a unit lozenge
in the broken line representation, or
$\pi'\subset \pi$, namely the broken line representation of 
$\pi'$ lies below that of $\pi$.

\noindent For illustration, when $k=3$, $n=2$ and $i=1$, we have $e_1 |\pi_0\rangle=|\pi_f\rangle+|\pi_2\rangle$ 
and $e_1|\pi_1\rangle=|\pi_3\rangle$
while $(e_1-\tau)$ annihilates $|\pi_2\rangle,|\pi_3\rangle,|\pi_f\rangle$, for the paths of Eq.~\fivepaths. 

Let us finally mention the rotation $\sigma$ acting on paths as follows. Given a path $\pi$, we record 
its last passages on the walls of the Weyl chamber, namely at points with $\lambda_m=0$, for
$m=1,2,\ldots,k-1$. The steps taken from those points are of the type $1,2,\ldots,k-1$ respectively. 
The rotated path $\sigma \pi$
is obtained by first "rotating" these steps, namely transforming them into steps of type $2,3,\ldots,k$
respectively, while all the other steps are unchanged,
then deleting the last step $\pi_N=k$, and finally adding a first step $1$. 
With this definition, we have the rotational invariance: $C_{i+1,\sigma\pi,\sigma\pi'}=C_{i,\pi,\pi'}$,
for $i=1,2,\ldots,N-2$, which may be recast into $\sigma e_i=e_{i+1}\sigma$, and thus allows for defining
an extra generator $e_N=\sigma e_{N-1} \sigma^{-1}=\sigma^{-1}e_1\sigma$.
As an example, the paths of Eq.~\fivepaths\ form two cycles under the rotation
$\sigma$, namely $\pi_0\to \pi_3\to\pi_0$ and $\pi_1\to \pi_2\to\pi_f\to\pi_1$.

\newsec{The quantum Knizhnik--Zamolodchikov equation}
The quantum Knizhnik--Zamolodchikov ($q$KZ) equation \FR\ is a linear difference equation satisfied by matrix elements of
intertwiners of highest weight representations of affine quantum groups (here, $U_q(\widehat{\goth{sl}(k)})$).
It can be reformulated as
the following equivalent set of conditions:
\eqna\amount
$$\eqalignno{
\tau_i \Psi(z_1,\ldots,z_N) &=\Rc_i(z_{i+1},z_{i}) \Psi(z_1,\ldots,z_N), \quad i=1,2,\ldots,N-1&\amount{a}\cr
\Psi(z_2,z_3,\ldots,z_N,s\, z_1)&=c\,\sigma^{-1} \Psi(z_1,\ldots,z_N)&\amount{b}\cr
}$$
where $\Psi=\sum_\pi\Psi_\pi |\pi\rangle$ is a vector in the path representation defined above. 
In \amount{a},
$\tau_i$ acts on functions of the $z$'s by interchanging $z_i$ and $z_{i+1}$. 
In \amount{b} 
$\sigma$ is the rotation operator: $(\sigma^{-1}\Psi)_\pi=\Psi_{\sigma\pi}$; $c$ is an
irrelevant constant which can be absorbed by homogeneity; $s$ determines
the {\it level}\/ $l$ of the $q$KZ equation via
$s=q^{2(k+l)}$, where $k$ plays here the role of dual Coxeter number. Note that instead
of imposing Eq.~\amount{b}, one could consider only
the Eqs.~\amount{a} but for any $i$, with the implicit shifted periodic boundary conditions $z_{i+N}=s\,z_i$,
and the various $\check R_i$ related to each other by conjugation by $\sigma$.

We introduce the operators $t_i$, acting locally on the variables
$z_i$ and $z_{i+1}$ of functions $f$ of $z_1,z_2,\ldots,z_N$ via
\eqn\genti{ t_i f= (q z_i-q^{-1} z_{i+1})\partial_i f }
where 
$\partial_i = {1\over z_{i+1}-z_{i}}(\tau_i-1)$
is the divided difference operator.\foot{The more
standard generator $T_i=q(q-t_i)$ of the Hecke algebra (which satisfies
the braid relations) is nothing but the Lusztig operator \Lu.}
Eq.~\amount{a} is equivalent to $t_i\Psi=(e_i-\tau)\Psi$, and indeed
one can check that $t_i+\tau$, or equivalently $-t_i$, 
satisfy by construction the Hecke algebra relations \relahec.
Decomposing $\Psi=\sum_\pi \Psi_\pi |\pi\rangle$
in the path representation
basis leads to
\eqn\master{t_i \Psi_\pi =\sum_{\pi'\neq \pi\atop \pi\in e_i\pi'} \Psi_{\pi'},\qquad i=1,2,\ldots,N-1 }
where the notation $\pi\in e_i \pi'$ simply means that we select the $\pi'$ such
that $C_{i,\pi',\pi}=1$. 

\newsec{Minimal polynomial solutions}
Looking for polynomial solutions of minimal degree to the set of equations \amount{},
we have found that 
\eqn\shiftc{ s=q^{2(k+1)}\qquad  {\rm and} \qquad c=\big((-1)^k q^{k+1}\big)^{n-1}\ ,}
so that these are solutions at level 1, and that 
\eqn\fundak{\Psi_{\pi_0}=\prod_{m=1}^k\ \prod_{1+(m-1)n\leq i<j\leq mn}(qz_i-q^{-1}z_j) }
To see why, first consider the equations \master, and set $z_{i+1}=q^2 z_i$: this
implies $e_i\Psi =\tau\Psi$, hence if $\tau\neq 0$ (i.e. $q^2\neq -1$), then
all the components $\Psi_\pi$ where $\pi$ is not in the image of some $\pi'$ under $e_i$
must vanish. Thanks to property {\bf (P3)}, we see that
\eqn\vani{ \Psi_\pi\vert_{z_{i+1}=q^2z_i}=0\qquad {\rm if}\quad \pi_i\geq \pi_{i+1}}
By appropriate iterations, this is easily extended to concave portions of $\pi$, namely such that 
$\pi_i\geq\pi_{i+1}\geq\cdots\geq\pi_{i+j}$, for which $\Psi_\pi$ vanishes at
$z_l=q^2z_k$, for any pair $k,l$ such that $i\leq k<l\leq j$. Henceforth, as $\pi_0$ has
$k$ flat portions $\pi_{(m-1)n+1}=\pi_{(m-1)n+2}=\cdots=\pi_{mn}=m$, $m=1,2,\ldots,k$, 
separated by convex points we find
that $\Psi_{\pi_0}$ must factor out the expression \fundak, which is the minimal realization of this
property. Alternatively, this may be recast into the highest weight condition that
$(t_i+\tau)\Psi_{\pi_0}=0$ for all $i$ not multiple of $n$, and \fundak\ is the polynomial of smallest
degree satisfying it.
Note that once $\Psi_{\pi_0}$ is fixed, all the other components of $\Psi$ are determined
by the equations \master, and therefore the cyclicity condition \amount{b} is automatically
satisfied, with the values of $c_N$ and $s$ \shiftc\ fixed by compatibility. This is a consequence of
the property {\bf(P4)}, which allows to express any $\Psi_\pi$ with $\pi_i<\pi_{i+1}$ only
in terms of $\Psi_{\pi'}$'s such that $\pi\subset \pi'$ in the above sense, henceforth in a triangular way w.r.t.
inclusion of paths.

Another consequence of this property is that if we pick say $z_m=z$, $z_{m+1}=q^2 z$, 
\dots, $z_{m+k-1}=q^{2(k-1)}z$
for $m+k\leq N$, then the only possibly non-vanishing components $\Psi_\pi$ of $\Psi$ are those 
having the convex sequence $\pi_m=1$, $\pi_{m+1}=2$, ... $\pi_{m+k-1}=k$. Let $\varphi_{m,m+k-1}$ denote
the embedding of $SU(k)$ paths of length $(n-1)k$ into those of length $nk$ obtained by inserting
a convex sequence of $k$ steps $1,2,\ldots,k$ between the $m-1$-th and $m$-th steps. Then we have
the following recursion relation:
\eqnn\recukembed
$$\eqalignno{
&\Psi_{\varphi_{m,m+k-1}(\pi)}(z_1,\ldots,z_N)\vert_{z_{m+k-1}=q^2z_{m+k-2}=...=q^{2(k-1)}z_m}&\recukembed\cr
&=C\left(\prod_{i=1}^{m-1}(qz_i-q^{-1}z_m)\prod_{i=m+k}^N(qz_{m+k-1}-q^{-1}z_i) \right)
\Psi_\pi(z_1,\ldots,z_{m-1},z_{m+k},\ldots,z_N)\cr}$$
for some constant $C$. This is readily obtained by first noting
that the equations \master\ are still satisfied by the l.h.s.\ of \recukembed\ for $i=1,2,\ldots,m-2$
and $i=m+k,m+k+1,\ldots,N$, while the prefactor in the r.h.s.\ remains unchanged. Moreover, 
expressing the interchange of
$z_{m-1}$ and $z_{m+k}$ in the l.h.s. as suitable successive actions of $\Rc$ matrices yields the
missing equation.

An alternative 
characterization of this polynomial solution is that all components
of $\Psi$ vanish whenever we restrict to 
\eqn\vani{z_{i_1}=z, \quad z_{i_2}=q^2z,\quad  ..., \quad z_{i_{k+1}}=q^{2k} z}
for some 
ordered $(k+1)$-uple $i_1<i_2<...<i_{k+1}$. 
This generalizes the observation of Pasquier
\Pas\ to the $SU(k)$ case and allows to interpret our solution as some $q$-deformed fractional quantum Hall 
effect wave functions, with filling factor $\nu=1/k$, bound to vanish whenever $k+1$ particles 
come into contact, up to shifts of $q^{2j}$, the so-called ``quantum incompressibility" condition.

Note that level 1 highest weight representations of $U_q(\widehat{\goth{sl}(k)})$
have been extensively studied in the literature; in particular,
$q$-bosonization techniques lead to integral formulae for solutions of level 1 $q$KZ equation, 
see e.g.~\JM, but they are usually expressed in the spin basis.

\newsec{Generalized Razumov--Stroganov sum rules and generalized ASM numbers}
To derive a sum rule for the components of $\Psi$, we introduce a covector $v$ such that
\eqn\vecond{
v\, e_i=\tau\, v, \quad i=1, 2,\ldots, N-1\qquad{\rm and}\qquad
v\, \sigma=v
}
Expressing $vY_k(e_i,...,e_{i+k-1})=0$, we find that $U_1(\tau)U_2(\tau)\cdots U_k(\tau)=0$ 
and further demanding that $v$ have
positive entries fixes
\eqn\valtau{ q=-e^{i\pi\over k+1} \qquad {\it i.e.} \qquad \tau=2\cos\left({\pi\over k+1}\right) }
Note that Eqs.~\vecond\ and the above path representation allow for writing a manifestly
positive formula for the entries of $v$ only in terms of Chebyshev polynomials, and we choose
the normalization $v_{\pi_f}=1$. For illustration, for $k=3$, $n=2$, $v$ is indexed
by the paths \fivepaths, and we have $v_{\pi_f}=1$, $v_{\pi_3}=U_1=\sqrt{2}$, $v_{\pi_2}=v_{\pi_1}=U_2=1$
and $v_{\pi_0}=U_1U_2=\sqrt{2}$.
The covector $v$ clearly satisfies $v\cR_{i,i+1}=v$ for $i=1,2,\ldots,N-1$ from the
explicit form \trigosol. We deduce that the quantity
$v\cdot \Psi$ is invariant under the action of $\tau_i$, $i=1,2,\ldots,N-1$, as a direct consequence
of \amount{a} and of the first line of \vecond, hence is
fully symmetric in the $z_i$'s. This symmetry is compatible with the cyclic relation
\amount{b} as $c=s=1$ from \valtau. Note that in the spin basis, $v\cdot\Psi$ is nothing but the sum of
all components of $\Psi$.

An important remark is in order: as we have $c=s=1$, $\Psi$ is actually the suitably normalized 
groundstate vector of the fully inhomogeneous $A_{k-1}$ IRF model
on a semi-infinite cylinder of perimeter $N$, defined via its (periodic) transfer matrix acting on 
the path basis states. 
Indeed, this transfer matrix is readily seen
to be (i) intertwined by the matrices $\Rc_{i,i+1}(z_{i+1},z_i)$ and (ii) cyclically symmetric under a rotation
by one step along the boundary of the half-cylinder, hence if $\Psi$ denotes the (Perron--Frobenius)
groundstate eigenvector of this transfer matrix, then the equations \amount{} follow, with $c=s=1$.
These models reduce to the half-cylinder $O(1)$ loop model for $k=2$, upon identifying the ``Dyck path"
basis with that of link patterns. In that case, the covector $v$ is simply $(1,1,\ldots,1)$, and
gives rise to the Razumov--Stroganov sum rule for $v\cdot\Psi$, proved in \DFZJ.
For general $k$, the corresponding sum rule reads:
\eqn\sumrul{ i^{kn(n-1)/2}\,  v\cdot \Psi = s_Y(z_1,\ldots ,z_N) }
where $i=\sqrt{-1}$, $s$ is a Schur function, and $Y$ is the Young diagram with $k$ rows of $(n-1)$ boxes,
$k$ rows of $(n-2)$ boxes, \dots, $k$ rows of $1$ box.
Eq.~\sumrul\ is proved as follows: by construction, $v\cdot \Psi$ is a symmetric polynomial, of total degree
$kn(n-1)/2$, and partial $n-1$ in each $z_i$, which moreover satisfies
recursion relations inherited from \recukembed, upon noting that
$v_{\varphi_{m,m+k-1}(\pi)}/v_{\pi}$ is independent of the path $\pi$. The Schur function on the r.h.s. is the unique 
symmetric polynomial of the $z$'s with the correct degree, and subject to these recursion relations
(or alternatively to the vanishing condition \vani). The
remaining global normalization is fixed by induction, by comparing both sides at $n=1$.

Using the explicit definition of Schur functions, we get the following homogeneous limits,
when all $z_i$'s tend to 1:
\eqn\genasm{
i^{kn(n-1)/2}\, v\cdot \Psi_{\rm Hom} =s_Y(1,1,\ldots,1)
=(k+1)^{n(n-1)/2} {\prod_{i=1}^{k-1}\prod_{j=0}^{n-1}((k+1)j+i)!\over
\prod_{j=0}^{n(k-1)-1} (n+j)! } }
where the quantities
\eqn\defasmgen{ A_n^{(k)} = {\prod_{i=1}^{k-1}\prod_{j=0}^{n-1}((k+1)j+i)!\over
\prod_{j=0}^{n(k-1)-1} (n+j)! } }
are integers that generalize the numbers of alternating sign matrices of size $n\times n$, recovered for
$k=2$. Note also that $A_2^{(k)}=c_k$, the $k$-th Catalan number.

It would be extremely interesting to find a combinatorial interpretation for these integer numbers, possibly
in terms of ``domain wall boundary condition" partition functions of the associated $SU(k)$ vertex model.
Note however that as opposed to the $SU(2)$ case, the components of $\Psi$ are no longer integers.

\newsec{Rational limit and geometry: extended Joseph polynomials}
For generic $q$ one could hope the polynomials defined above to be geometrically interpreted in terms of K-theory;
this is however beyond the scope of the present letter, and we only consider here
the rational limit, which is obtained by substituting
\eqn\jolim{ q=-e^{-\epsilon a/2},\qquad z_i=e^{-\epsilon w_i}, \ i=1,2,\ldots,N} 
and expanding to first non-trivial order in $\epsilon$ as it goes to zero.

The $q$KZ equation becomes the ``rational $q$KZ equation'': it has the same form as before, 
but the $R$-matrix is now 
the rational solution of YBE related to quotients of the symmetric group ${\cal S}_N$.

In this limit, 
the $q$KZ polynomial solutions at level 1 remain polynomials in $a,w_1,\ldots,w_N$.
They turn out to have a remarkable algebro-geometric interpretation, in the same 
spirit as \KZJ: they are ``extended'' Joseph
polynomials \DFKZJ\ associated to the Young diagram of rectangular shape $k\times n$.
More precisely, consider the scheme of nilpotent
$N\times N$ complex upper triangular matrices $U$ which satisfy $U^k=0$. 
It is well-known that its irreducible components are indexed by Standard Young Tableaux (SYT) 
of rectangular shape $k\times n$; the latter are equivalent to paths in the Weyl chamber 
of $SU(k)$, according to the following rule: the numbers on the $i^{\rm th}$ row of the SYT 
record the positions of steps of type $i$ of the path.
To each path one can then associate the equivariant multiplicity \refs{\Ro,\Jo}\ 
(also called {\it multidegree}) 
of the corresponding component with respect to the action
of the torus $({\Bbb C}^\times)^{N+1}$, where a $N$-dimensional torus acts by conjugation of $U$ by 
diagonal matrices and the extra one-dimensional torus acts by an overall rescaling of $U$. 
This results in a set of homogeneous polynomials in $N+1$ variables $a,w_1,\ldots,w_N$, 
which turn out to coincide with the entries of $\Psi$ in the limit \jolim\ above.
We call these polynomials {\it extended Joseph polynomials} because the usual 
Joseph polynomials \Jo, defined without the rescaling action, correspond to $a=0$.
On the other hand, note that setting $w_i=0$ and $a=1$ yields the degrees of the components.

The fact that extended Joseph polynomials satisfy the limit \jolim\ of
Eqs.~\master, is nothing but Hotta's explicit construction \Ho\ of the 
Joseph/Springer
representation. This will be discussed in detail elsewhere \DFKZJ. 
As a corollary, note that at $a=0$, the (usual) Joseph polynomials 
satisfy the (usual) level 1 Knizhnik--Zamolodchikov equation \KZ, which takes the form:
\eqn\KZ{(k+1) {\der\over\der w_i} \Psi=\sum_{j(\ne  i)} {s_{i,j}+1\over w_i-w_j} \Psi}
where $s_{i,j}$ is the transposition $(ij)$ in ${\cal S}_N$ (explicitly,
for $i<j$, $s_{j,i}=s_{i,j}=s_is_{i+1}\cdots s_{j-2} s_{j-1} s_{j-2}\cdots s_{i+1}s_i$ with
$s_i=1-e_i$).
The solutions of these equations are well-known in the spin basis, 
indexed by 
sequences $(\alpha_1,\ldots,\alpha_N)$ in which every number from $1$ to $k$
occurs $n$ times.
They are simple products of Vandermonde determinants:
$\Psi_{\alpha_1,\ldots,\alpha_N}(0,w_1,\ldots,w_N)=\prod_{\beta=1}^k \ 
\prod_{i,j\in A_\beta,i<j}(w_i-w_j)$,
where $A_\beta=\{\, i: \alpha_i = \beta \,\}$. 

%
A last remark is in order. Throughout this note, we have restricted ourselves for simplicity
to systems of size $N=kn$ multiple of $k$. However, we may obtain solutions of the $q$KZ 
equation
for arbitrary size say $N=kn-j$ from that of size $N=nk$ by letting successively 
$z_{kn}\to 0$, $z_{kn-1}\to 0$, \dots, $z_{kn-j+1}\to 0$ in $\Psi$. This immediately yields other
sum rules of the form \sumrul, but with a Schur function for a truncated Young
diagram, with its $j$ first rows deleted. Similarly, we have access to the multidegrees of
the scheme of nilpotent upper triangular matrices of arbitrary size as well.

\centerline{\bf Acknowledgments}
PZJ would like to thank H.~Boos, N.~Reshetikhin, F.~Smirnov for enlightening discussions, 
and the organizers of the Bad Honnef summer school on Representation Theory where some of these took place.
PDF thanks V.~Pasquier for useful conversations.
The authors thank A.~Knutson and J.-B.~Zuber for their help in the framework of parallel
collaborations, and acknowledge the support of the Geocomp project (ACI Masse de Donn\'ees)
and of the European network ``ENIGMA", grant MRT-CT-2004-5652.

\listrefs
\end